# Effect of high pressure synthesis conditions on the formation of high entropy oxides

**Authors:**

Solveig Stubmo Aamlid,[1] Minu Kim,[2] Mario U. González-Rivas,[1,3] Mohamed Oudah,[1] Hidenori Takagi,[2,4,5] Alannah M. Hallas[1,3,6,*]

*alannah.hallas@ubc.ca

**Affiliations:**

1. Stewart Blusson Quantum Matter Institute, University of British Columbia, Vancouver, BC V6T 1Z4, Canada

2. Max Planck Institute for Solid State Research, Heisenbergstrasse 1, 70569 Stuttgart, Germany

3. Department of Physics & Astronomy, University of British Columbia, Vancouver, BC V6T 1Z1, Canada

4. Department of Physics, University of Tokyo, Bunkyo-ku, Hongo 7-3-1, Tokyo 113-0033, Japan

5. Institute for Functional Matter and Quantum Technologies, University of Stuttgart, 70550 Stuttgart, Germany

6. Canadian Institute for Advanced Research, Toronto, ON M5G 1M1, Canada

**Abstract:** High entropy materials are often entropy stabilized, meaning that the configurational entropy from multiple elements sharing a single lattice site stabilizes the structure. In this work, we study how high-pressure synthesis conditions can stabilize or destabilize a high entropy oxide (HEO). We study the high-pressure and high-temperature phase equilibria of two well-known families of HEOs: the rock-salt structured compound $(Mg,Co,Ni,Cu,Zn)O$ including some cation substitutions and the spinel structured $(Cr,Mn,Fe,Co,Ni)_3O_4$. Syntheses were performed at various temperatures, pressures, and oxygen activity levels resulting in dramatically different synthesis outcomes. In particular, in the rock salt HEO we observe the competing tenorite and wurtzite phases and the possible formation of a layered rock salt phase, while the spinel HEO is highly susceptible to decomposition into a mixture of rock-salt and corundum phases. At the highest tested pressures, 15 GPa, we discover the transformation of the spinel HEO into a metastable modified ludwigite-type structure with nominal formula $(Cr,Mn,Fe,Co,Ni)_4O_5$. The relationship between the synthesis conditions and the final reaction product is not straight-forward. Nonetheless, we conclude that high-pressure conditions provide an important opportunity to synthesize high entropy phases that cannot be formed any other way.

High entropy oxides (HEOs) are oxides where one crystallographic site is shared by five cations in roughly equimolar ratios[1–4]. In order for an HEO to form, there are a number of conditions that must be met: each binary constituent oxide must have reasonable transformation enthalpies from their ground state structure to the shared structure, there cannot be any competing cation ordered phases, and the cations must also be similarly sized so the oxygen lattice can mediate the strain interactions[5,6]. High entropy phases are sometimes entropy stabilized, meaning that the configurational entropy from the cation disorder stabilizes the resulting phase rather than enthalpy. Entropy stabilized HEOs can therefore be regarded as metastable phases[7].

Temperature alone provides a very narrow parameter space in which to explore the formation of HEOs. In this regard, there are two main advantages of applying high pressure synthesis to HEOs. The first is that it enables control over the pressure-volume (pV) term in the thermodynamic free energy, providing a switch from a one-dimensional parameter space (temperature) to a two-dimensional one (temperature and pressure). While the prediction of high-pressure phases is not trivial, generally compounds with higher densities and coordination numbers are selected at elevated pressures. The second advantage is that the pressurized vessel provides a sealed environment, which allows for control over oxidation states through the inclusion of an oxidizing reagent. The applied pressure and oxygen partial pressure acts as two somewhat related pressure axes. Alternatively, lower oxidation states can be achieved by selecting appropriate precursors. As in solid-state synthesis, high-pressure reactions provide good control over stoichiometry since there is no preferential precipitation, but with faster diffusion, which is beneficial to achieving the ideal mixing of an HEO. Some reports on high-pressure properties and syntheses of HEOs exist,[8–10] but the method has not yet been widely applied. Therefore, the use of high pressure to form new phases or control oxidation states in HEOs may pave the way to accessing structures and compositions that cannot be made any other way.

In this work, we explore the use of high-pressure synthesis to expand theas phase space for high entropy oxides. We take two well characterized HEOs, the rock salt (Mg,Co,Ni,Cu,Zn)O and spinel (Cr,Mn,Fe,Co,Ni)$_3$O$_4$, as our test systems[1,11]. We set out to demonstrate a proof-of-concept of stabilization and destabilization of HEOs under high pressure, and to stabilize HEOs with unusual oxidation states. We uncover a complex interdependence of temperature, pressure, and oxygen partial pressure in this phase space that strongly modifies the structure selection. At the highest tested pressures, we discover a new metastable polymorph of the spinel HEO with a modified ludwigite structure.

The high-pressure synthesis of the HEO spinel (Cr,Mn,Fe,Co,Ni)$_3$O$_4$ samples used a pre-reacted and unannealed powder from a glycine-nitrate spontaneous combustion synthesis following a previously reported method[12]. An XRD pattern of the precursor powder is shown in Figure S1, it has reacted to form a spinel phase and displays broad peaks due to the nanosize of the powder. When indicated, 4 or 10% KClO$_4$ oxidizer by weight, was ground into the powder. The high-pressure synthesis of the rock-salt compounds followed a conventional mixed oxide route where all the precursors were already in the nominal cation-oxygen 1:1 stoichiometry, but not pre-reacted before pressurization. In all cases, the mixed powder was loaded into a Pt capsule and compressed to the synthesis pressure. The sample was heated to the reaction temperature and held for 1 hour before thermal quenching down to room temperature followed by slow decompression to ambient pressure. A full overview of chemical compositions and reaction pressures and temperatures are summarized in Tables I and ii. A Walker-type multi-anvil module was used for all the spinel samples, the 15 GPa prototype rock-salt sample, and the cation substituted rock-salt samples, and a belt module was used for the remaining prototype rock-salt samples. The resulting products were taken out of the capsule and ground to a powder. If applicable, any remaining KCl salt from the oxidizer was washed away using distilled water and ethanol.

Phase composition of the synthesized materials was assessed using powder x-ray diffraction (XRD). These measurements were performed on a Bruker D8 Advance diffractometer using a monochromated Cu x-ray source in Bragg-Brentano geometry. Phase matching was done using DIFFRAC.EVA and the Crystallographic Open Database (COD)[13]. The COD catalog entries used for phase matching can be found in Table SI in the Supplementary Materials. The Rietveld refinement of the modified ludwigite phase was performed in TOPAS[14] using ICSD-266256 as a starting point[15,16]. Since the cation distribution between the three sites is unknown and unresolvable to conventional x-rays, pure Fe was used to model the cation mixture. Lattice parameters and atomic positions were refined, while thermal parameters were not. For the peak shapes, instrumental parameters and Lorentzian and Gaussian strain were refined. For the calculation of the full width at half maximum (FWHM) for the rock-salt phases, individual peaks were modelled using peak position, intensity, and a Lorentzian strain broadening as free parameters. The temperature-dependent XRD measurements were performed on the same instrument using a Bruker MTC Furnace stage under low pressure (1 mbar), resulting in mildly reducing conditions. Samples were mounted on an alumina holder. XRD patterns were collected at 30 °C and in 100 °C steps on warming from 100 °C up to 900 °C, with a 50 °C/min ramp rate, a 15 min stabilization dwell time, and a 60 min collection time. Throughout this paper, multiple elements in the same parenthesis signify that these elements are present in equimolar ratios.

In order to establish a baseline for our high-pressure experiments, we begin our study by considering the stabilizing or destabilizing effect of temperature alone on the stability of the rock salt and spinel HEOs. Both the spinel and rock-salt samples used in the temperature-dependent XRD measurements were produced by combustion of an appropriate glycine-nitrate solution and annealed to improve crystallinity. Their XRD patterns are shown in Figure S2. The prototypical cubic rock salt HEO (Mg,Co,Ni,Cu,Zn)O has a close-packed structure of edge-sharing octahedra, as shown in **Fig. 1(a)**. On heating in vacuum at this heating rate, the rock salt structure is found to start decomposing at 800 °C when secondary phases start to grow (marked with stars in **Fig. 1(b)**). This is consistent with experiments performed in air on both warming and cooling which show that the rock salt is susceptible to both de-mixing and tetragonal distortions if kept in thermal equilibrium between 600 and 800 °C, depending on equilibration time [17,18]. The spinel HEO $(Cr,Mn,Fe,Co,Ni)_3O_4$ shown in **Fig. 1(c)** has two distinct cation sublattices with a mixture of divalent and trivalent cations that have strong preferences for the octahedral and tetrahedral sites[11,19]. Temperature dependent XRD shows significant decomposition for the spinel HEO that manifests in peak broadening from 600 °C and the appearance of secondary phases from 800 °C (**Fig. 1(d)**). Both samples display conventional thermal expansion. The decomposition of both these high entropy phases below their synthesis temperature is consistent with metastability and hence entropy stabilization.

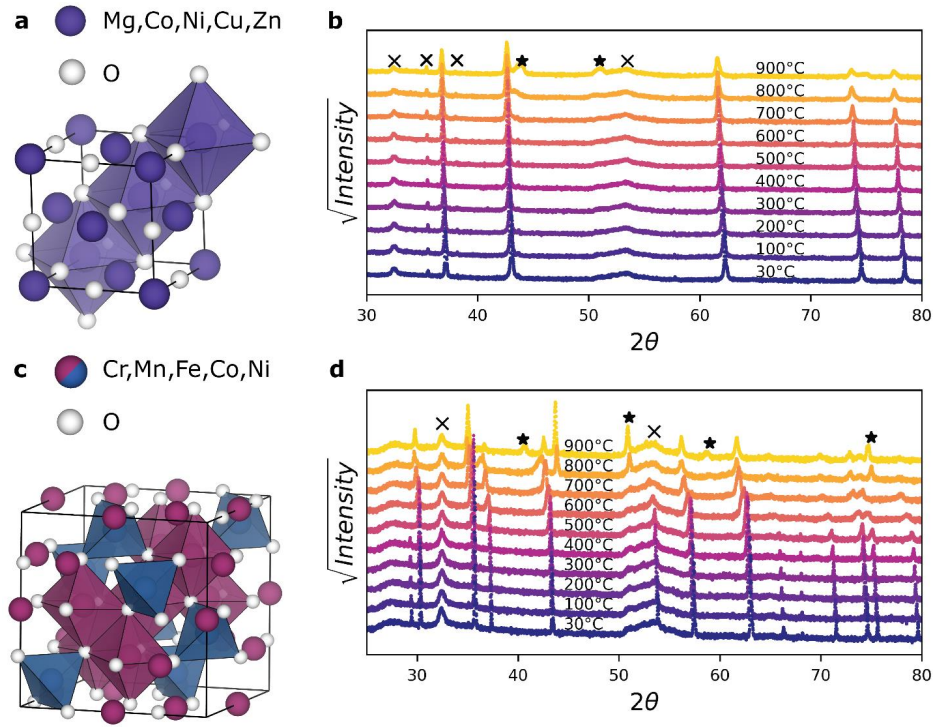

*Figure 1* **a** The rock-salt crystal structure, **b** the temperature-dependent x-ray diffractograms of the rock-salt sample with x marking the background from the sample stage and stars marking the reflections from dissolution impurities. **c** The spinel crystal structure, **d** the temperature-dependent x-ray diffractograms of the spinel sample with equivalent markers.

Moving now into the two-dimensional pressure-temperature phase space, we explore the stability of the spinel HEO under a wide range of pressures and oxidizer concentrations. The synthesis parameters and resulting phase compositions for the $(Cr,Mn,Fe,Co,Ni)_3O_4$ spinel samples are displayed in **Table I** and **Fig. 2(a)** shows the phase matched x-ray diffractograms. The outcome of the phase identification based on the applied pressure and oxidizer content is shown schematically in **Fig. 2(b)**. The spinel structure is unstable towards rock salt and corundum phases under a wide range of conditions. If we first consider the three 10 GPa syntheses with increasing oxidizer content, we observe that too little oxidizer in the 0% case, leads to some formation of the reduced divalent rock-salt structure, while too much oxidizer in the 10% case results in the formation of the oxidized trivalent corundum structure. At 10 GPa, around 4% oxidizer is necessary to form a phase pure spinel sample, which we ascribe to a slight oxygen deficiency in the starting material, as spinels are prone to oxygen vacancies[20]. Notably, at 5 GPa and 4% oxidizer, the resulting phase mixture is still slightly reduced relative to the target spinel phase, as the oxidizer content together with the temperature and applied pressure determines the oxygen activity level. The 1 GPa and 10% oxidizer synthesis (which is also performed at a lower temperature) again highlights the complicated interdependence of the pressure, temperature, and oxidizer content in this composition space, as no corundum phase is observed at this relatively lower temperature and pressure.

*Table I* Overview of the experimental conditions used during the high-pressure synthesis attempts for the spinel structured HEO.

| Label | Pressure (GPa) | Temperature (°C) | Oxidizer, $KClO_4$ (wt. %) | Phase composition (from XRD) |
| --- | --- | --- | --- | --- |
| VII | 15 | 1000 | - | Modified ludwigite + minor impurity |
| VI | 15 | 1000 | 10 | Corundum + rock-salt + minor impurity |
| V | 10 | 1000 | 10 | Spinel + corundum |
| IV | 10 | 1000 | 4 | Spinel |
| III | 10 | 1000 | - | Spinel + rock-salt |
| II | 5 | 1000 | 4 | Spinel + rock-salt |
| I | 1 | 850 | 10 | Spinel |

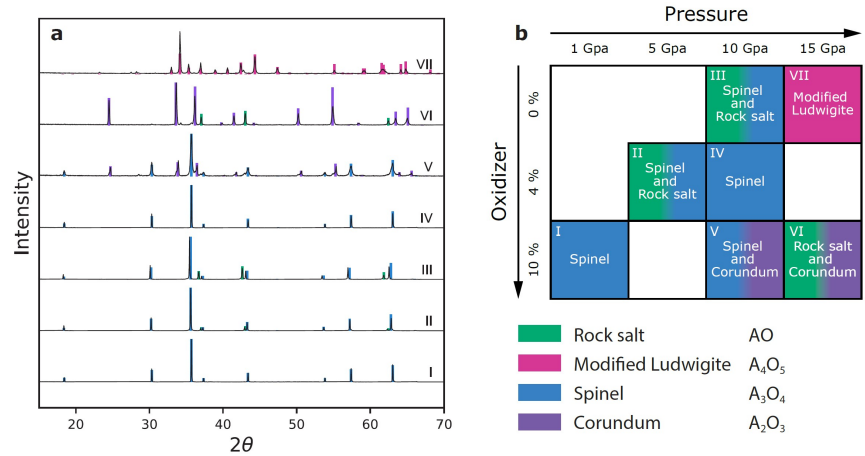

*Figure 2* **a** X-ray diffractograms of the spinel samples synthesized under different temperatures, applied pressures, and oxygen partial pressure conditions. **b** A summary of the resulting phases as a function of the applied pressure and oxidizer content.

At the highest applied pressures, 15 GPa, and without oxidizer we observe a complete transformation of the spinel HEO into a metastable modified ludwigite structure. The nominal stoichiometry of this orthorhombic phase is $(Cr,Mn,Fe,Co,Ni)_4O_5$ and it contains three distinct cation environments, as shown in **Fig. 3(a)**. A Rietveld refinement of this structure (space group *Cmcm*, #63) is presented in **Fig. 3(b)**, the refined parameters can be found in Table SII in the Supplementary Materials. The most common high-pressure, high-density post-spinel phases are the $CaFe_2O_4$, $CaTi_2O_4$, and $CaMn_2O_4$ types, all of which are orthorhombic structures that maintain the $AB_2O_4$ spinel stoichiometry[21]. In contrast, modified ludwigite structures are sometimes found at lower pressures than the other post-spinel phases and due to the slightly reduced oxidation state of the cations often found coinciding with the comparatively oxidized corundum structures[22]. The modified ludwigite structure found here consists of edge- and corner-sharing octahedra and trigonal prisms, whereas the spinel structure is comprised of edge- and corner-sharing octahedra and tetrahedra. The difference in the crystal field effects between a tetrahedra and an octahedra is much larger than the difference between a trigonal prism and an octahedra[23]. At first glance, the strong site selectivity previously observed in the spinel phase[11,19] could be expected to practically vanish in the modified ludwigite-type structure, resulting in a more configurationally disordered, truly random high entropy compound. However, site selectivity even between the two inequivalent octahedral sites has been observed in a similar compound, so more investigations are needed to conclude in this matter[16].

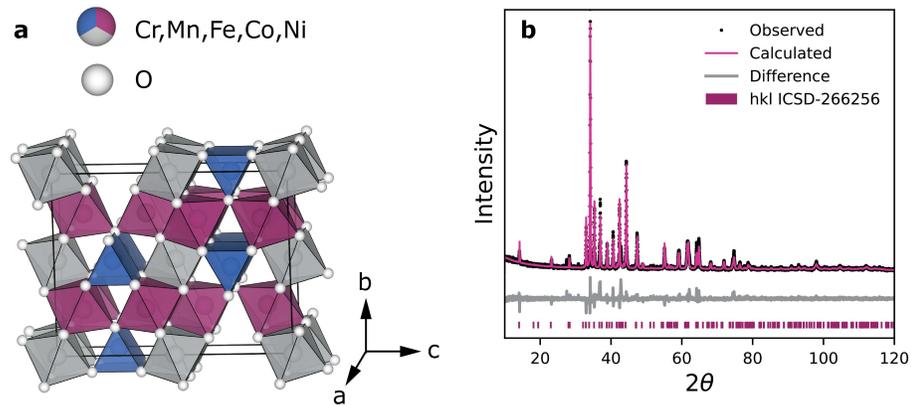

*Figure 3 **a** The high-pressure modified ludwigite structure **b** The Rietveld refinement of this phase, a few unindexed peaks signify some degree of impurity.*

The synthesis at 15 GPa with 10% oxidizer shows that this modified ludwigite phase is fragile with respect to high oxygen partial pressures, as it does not form in this highly oxidizing environment. Instead, the predominant phase is an oxidized corundum, which has a uniform trivalent cation oxidation state. However, at least one element does not incorporate in the corundum structure and instead forms a rock-salt phase. Out of the five elements present, Ni is the least likely to exist in a trivalent state. It is possible that a phase pure HEO corundum could be achieved either through replacement of Ni with a different transition metal or further increasing the oxygen partial pressure. It is interesting to note that the spinel structure is completely destabilized at 15 GPa regardless of the presence of oxidizer although the resulting phase composition is vastly different depending on the oxidizer content.

We next consider the effect of high-pressure synthesis conditions on the prototype rock-salt HEO $(Mg,Co,Ni,Cu,Zn)O$[1]. In contrast to the spinel scenario above, here we seek to maintain the uniform divalent cation oxidation state so no oxidizer was included. Instead, syntheses at various temperatures and pressures were attempted with the goal of establishing phase boundaries at moderate pressure, then increase the pressure to look for changes in formation temperature. The synthesis procedures and resulting phase compositions in the rock-salt system are shown in Table II and the corresponding XRD patterns are shown in **Fig. 4(a)**. In this set of experiments, accurate temperature control was not achieved. While the set power-percentage (PP) delivered to the instrument was kept constant during synthesis, the temperature was observed to drop steadily through the 1 hour hold time. The temperature ranges and averages estimated from calibrations are shown in Table II.

Attempts to synthesize the rock salt HEO at 1.5 GPa with reaction temperatures below 800 °C show a mixture of phases, including tenorite (CuO) and wurtzite (ZnO). This result is consistent with the ambient pressure temperature-dependent XRD shown previously in **Fig. 1(b)**. At the same pressure, reactions performed above 800 °C lead to a single-phase rock salt. An interesting temperature dependence is uncovered in the 6 GPa growths where CuO and ZnO are both observed to phase separate at temperatures around 750 °C but a slight increase to around 800 °C leads ZnO to be incorporated while CuO remains in the tenorite structure. This is perhaps not surprising as ZnO in and of itself takes on the rock-salt structure around 6 GPa[24]. It is interesting to compare this 6 GPa synthesis at 800 °C with the one performed at 1.5 GPa and 830 °C, which yields phase pure rock salt. It is unclear whether it is the increased pressure or the small decrease in temperature that causes CuO to stay out of the common high entropy phase in the 6 GPa synthesis.

Another intriguing outcome for the rock salt HEO occurs at the highest tested pressure. The synthesis at 15 GPa resulted in a partial decomposition of the rock-salt phase to a new phase with (*hkl*) indices matching those of $CoO_2$ or $Mg(OH)_2$ (hexagonal space group *P*-3*m*1, #164). There is no surplus oxygen or hydrogen present during the synthesis so the stoichiometry should not change without severe degradation of the Pt capsule, which was not detected. The observation of this space group could be an effect of some kind of long-range ordering of the cations in layers (similar to how Li can intercalate between the layers in $CoO_2$ to produce a similar diffraction pattern albeit with different intensities). A theoretical work supports various orderings in the rock salt (Mg,Zn)O under high pressure[25], so this kind of behavior is not unheard of in related systems. In the future, it will be worthwhile to study the evolution of the rock salt HEO under even higher pressures.

Considering now only the rock salt syntheses that resulted in the highest phase purity products, we can see that there is a large and anisotropic peak broadening, as shown in **Fig. 4(b)**. Relative differences in intensity and FWHM of the (111) and (200) reflections in this rock-salt compound have previously been connected to the extent of the Jahn-Teller distortion[18]. Increased Cu content and slower cooling rates both lead to a larger disparity in these values as local Jahn-Teller distortions around the Cu cations are promoted. Comparitively, the cooling rate in the high-pressure experiments performed here is nearly instantaneous due to the good thermal contact for heat dissipation after the heater is turned off. Individual peak fitting was used to extract the FWHM of the four samples. The FWHM for the (111) and (200) reflections and the ratio between them (anisotropy) are shown in Table III. Comparing the two syntheses at 1.5 GPa, the synthesis at higher temperature has more narrow peak shapes overall, but the ratio between them is similar. The same can be said about the synthesis at similar temperatures but higher pressure. The degree of anisotropy is similar in these three cases, although largest in the 6 GPa case. The anisotropy in the strain broadening is most pronounced in the sample synthesized at 15 GPa. To summarize, all the samples display a large and anisotropic peak broadening, but the highest pressure or the associated phase separation does amplify the effect.

*Table II* Overview of the experimental conditions used during the high-pressure synthesis attempts for the rock salt structured HEO.

| Pressure (GPa) | Average temperature (°C) | Temperature range (°C) | PP (%) | Phase composition (from XRD) |
|---|---|---|---|---|
| 15 | 870 | N/A | N/A | Rock-salt and *P*-3*m*1 |
| 6 | 800 | 840-750 | 39.0 | Rock-salt and CuO |
| 6 | 750 | 770-730 | 38.5 | Rock-salt, CuO, and ZnO |
| 1.5 | 930 | 960-890 | 41.5 | Phase pure rock-salt |
| 1.5 | 830 | 860-810 | 39.0 | Phase pure rock-salt |
| 1.5 | 780 | 780 | 38.6 | Rock-salt, CuO, and ZnO |
| 1.5 | 750 | 760-740 | 38.0 | Phase mixture (some reaction) |
| 1.5 | 710 | 710-700 | 37.5 | Phase mixture |

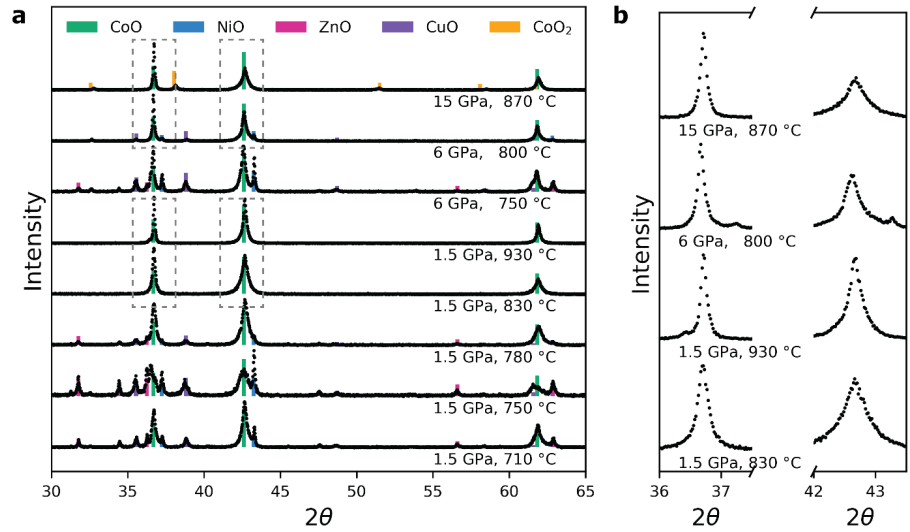

*Figure 4 **a** X-ray diffractograms of the prototype rock-salt under various temperature and pressure conditions **b** The 111 and 200 reflections of four selected rock-salt samples marked by the dashed boxes in **a** showing the anisotropic peak broadening.*

*Table III The FWHM and anisotropic peak broadening for the rock salt HEO synthesized with varying temperature and pressure.*

| Experimental conditions | FWHM 111 (degrees) | FWHM 200 (degrees) | FWHM200/FWHM111 anisotropic strain broadening |
|---|---|---|---|
| 15 GPa, 870 °C | 0.29 | 1.02 | 3.57 |
| 6 GPa, 800 °C | 0.36 | 0.86 | 2.37 |
| 1.5 GPa, 930 °C | 0.32 | 0.69 | 2.14 |
| 1.5 GPa, 830 °C | 0.57 | 1.18 | 2.07 |

Finally, several attempts were made to exploit high pressure conditions in order to facilitate the substitution of non-magnetic cations ($Zn^{2+}$ and $Mg^{2+}$) with magnetic cations ($Mn^{2+}$ and $Fe^{2+}$) in the rock-salt HEO. The divalent oxidation states of Mn and Fe (as opposed to their trivalent oxidation states which are more thermodynamically favorable) makes the incorporation of these elements unfavorable under ambient conditions[26]. However, even under the controlled environment of a high-pressure synthesis, we find certain incompatibilities between constituents, resulting in the mixture of phases formed displayed in **Fig. 5**. The reaction conditions are 2.5 GPa and set temperatures between 950 and 1060 °C, as noted in the figure. When Mn and Cu coexist, we observe the formation of a delafossite-type structure (where Cu is monovalent) while the combination of Fe and Cu leads to the formation of an oxidized spinel phase and reduced Cu metal. Meanwhile, Zn appears to be incompatible with Mn due to the formation of a secondary phase that strangely matches a tetravalent $ZrO_2$ baddeleyite phase. From these observations we can infer that a compound consisting of Mg, Co, Ni, Mn, and Fe might be synthesizable as it avoids all unfavorable pairings. Interestingly, exactly this composition has been synthesized through an alternative route[27]. Mechanochemical synthesis has also been used to stabilize the divalent oxidation state of Fe and Mn in a rock salt HEO, where the aforementioned incompatibilities do not inhibit phase formation[28].

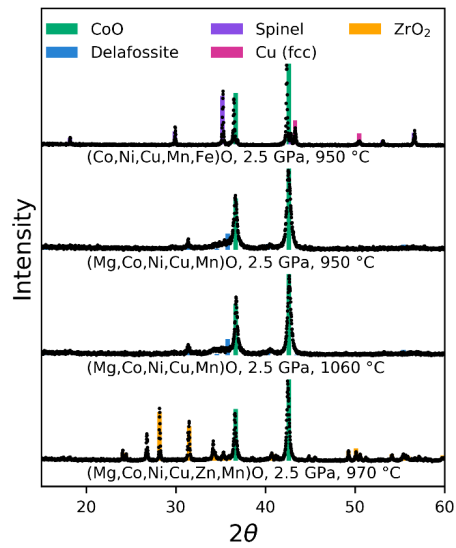

*Figure 5 X-ray diffractograms of the synthesis attempts at magnetic rock-salts with unstable oxidation states. None of the attempts result in a phase pure compound.*

In summary, our work highlights the complex interdependence of synthesis temperature, applied pressure, and oxygen partial pressure in the formation of high entropy phases. We show that high pressure can both stabilize and destabilize high entropy phases and can also stabilize high entropy materials with both higher and lower oxidation states. High-pressure synthesis is a worthwhile tool for materials discovery also in the high-entropy space although the relationship between the synthesis conditions and the final reaction product is not straightforward.

**Supplementary materials:** Supplementary materials contain Table SI with a list of phases from the Crystallographic Open Database, Table SII with the crystallographic information for the Rietveld refinement of the modified Ludwigite, Figure SI with the XRD of the pre-reacted and unannealed powder from the glycine-nitrate spontaneous combustion synthesis which was used in the high-pressure syntheses, and Figure SII with the XRD of the annealed powders used in the temperature-dependent X-ray diffraction study.

**Acknowledgements:** The authors acknowledge Graham McNally, Aleksandra Krajewska, Uwe Engelhardt, Frank Falkenberg for their support during the high-pressure synthesis experiments. Thanks to Joerg Rottler at UBC for helpful discussions. This research was undertaken thanks in part to funding from the Max Planck-UBC-UTokyo Centre for Quantum Materials and the Canada First Research Excellence Fund, Quantum Materials and Future Technologies Program. This work was also supported by the Natural Sciences and Engineering Research Council of Canada (NSERC), the CIFAR Azrieli Global Scholars program, and the Sloan Research Fellowships program. This work is funded in part by a QuantEmX grant from ICAM and the Gordon and Betty Moore Foundation through Grant GBMF9616 to Solveig Stubmo Aamlid.

**Conflict of interest:** The authors have no conflicts to disclose.

**Author contributions: Solveig Stubmo Aamlid**: conceptualization (equal), formal analysis (lead), investigation (equal), visualization (lead)**,** writing – original draft (lead), writing – review and editing (equal). **Minu Kim**: investigation (equal), resources (lead), writing – review and editing (equal). **Mario U. González-Rivas:** investigation (equal), writing – review and editing (equal). **Mohamed Oudah:** investigation (equal), writing – review and editing (equal). **Hidenori Takagi:** funding acquisition (equal), resources (equal), supervision (equal). **Alannah Hallas:** conceptualization (equal), funding acquisition (equal), supervision (equal), visualization (supporting), writing – original draft (supporting), writing – review and editing (lead).

**Data availability:** The data that support the findings of this study are available from the corresponding author upon reasonable request.